# Machine learning for evolutionary-based and physics-inspired protein design: current and future synergies.

Cyril Malbranke*[1,2], David Bikard[2], Simona Cocco[1], Rémi Monasson[1] and Jérôme Tubiana*[3]

1   Laboratory of Physics of the Ecole Normale Supérieure, PSL Research, CNRS UMR 8023, Sorbonne Université, Université de Paris, Paris, France
2   Institut Pasteur, Université Paris Cité, CNRS UMR 6047, Synthetic Biology, 75015 Paris, France
3   Blavatnik School of Computer Science, Tel Aviv University, Tel Aviv, Israel

* Correspondence: cyril.malbranke@phys.ens.fr, jertubiana@gmail.com

## Highlights

1. Machine learning methods for protein design are rapidly progressing along two parallel tracks: evolutionary-based and physics-inspired approaches.
2. Here, we recapitulate main progresses for both classes, and discuss their respective strengths and limitations.
3. We argue that both methods are highly complementary and discuss current and future synergistic approaches.

## Abstract

Computational protein design facilitates discovery of novel proteins with prescribed structure and functionality. Exciting designs were recently reported using novel data-driven methodologies that can be roughly divided into two categories: evolutionary-based and physics-inspired approaches. The former infer characteristic sequence features shared by sets of evolutionary-related proteins, such as conserved or coevolving positions, and recombine them to generate candidates with similar structure and function. The latter estimate key biochemical properties such as structure free energy, conformational entropy or binding affinities using machine learning surrogates, and optimize them to yield improved designs. Here, we review recent progress along both tracks, discuss their strengths and weaknesses, and highlight opportunities for synergistic approaches.

# Graphical Abstract

**I - Evolution-based approaches**

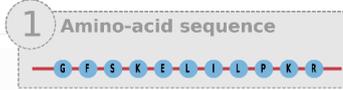
1. Amino-acid sequence

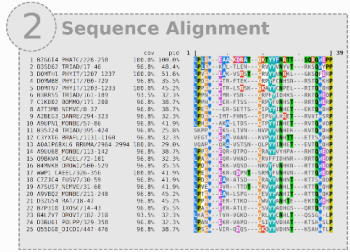
2. Sequence Alignment

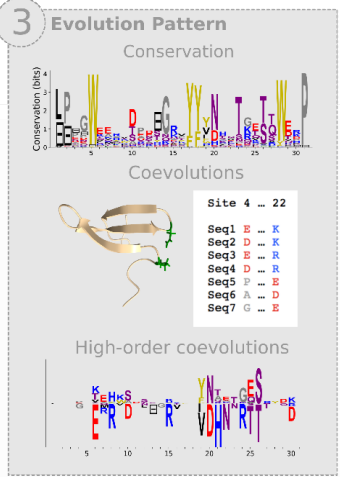
3. Evolution Pattern — Conservation, Coevolutions, High-order coevolutions

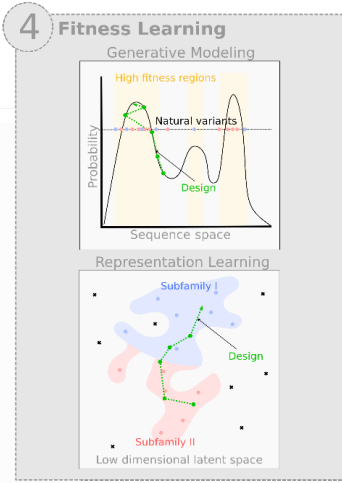
4. Fitness Learning — Generative Modeling, Representation Learning

**III - Synergistic approaches**

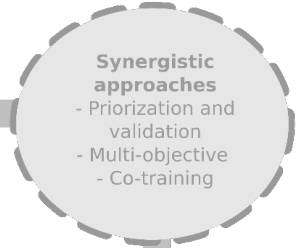

Synergistic approaches
- Priorization and validation
- Multi-objective
- Co-training

**II - Physics-grounded approaches**

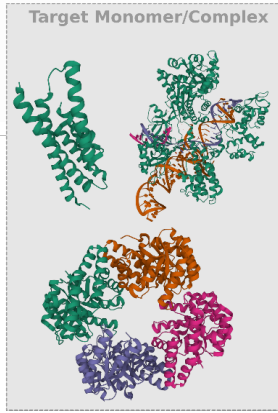
Target Monomer/Complex

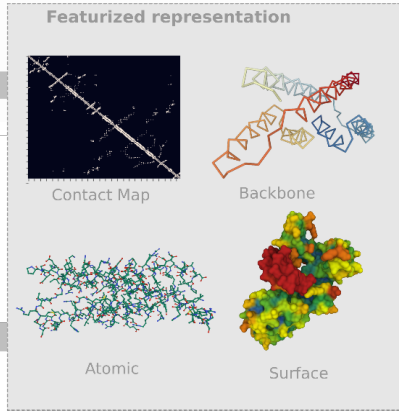
Featurized representation — Contact Map, Backbone, Atomic, Surface

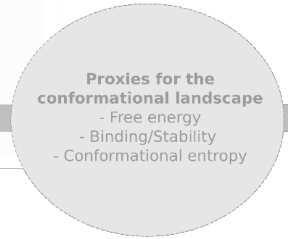
Proxies for the conformational landscape
- Free energy
- Binding/Stability
- Conformational entropy

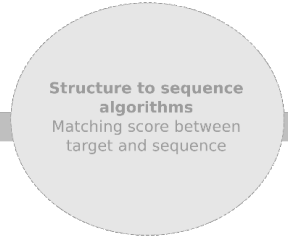
Structure to sequence algorithms
Matching score between target and sequence

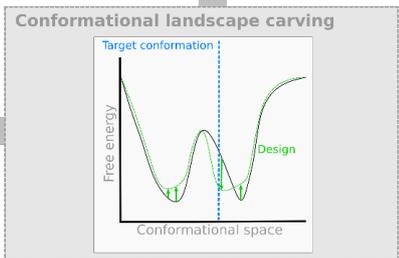
Conformational landscape carving

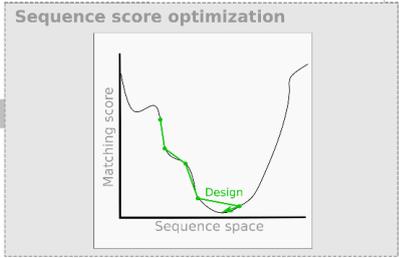
Sequence score optimization

# Introduction

Recent years have witnessed tremendous developments in computational protein design methodologies, following two parallel tracks: evolutionary-based design and physics-inspired design. The former have been propelled by i) increasingly fast sequencing and homology detection methodologies, allowing the constitution of large sequence databases such as Uniprot [1], structured into families of evolutionary-related proteins such as PFAM [2] and ii) novel unsupervised machine learning approaches for generative modeling of sequences (see review by Wu et al. [3]). The latter have been stirred by the emergence of a variety of deep learning models for predicting properties of proteins from sequence, structure or both (see review by Ovchinnikov and Huang [4]). Such prediction-based design protocols are highly appealing compared to traditional physics-based protein design protocols based on e.g. Rosetta or FoldX. Indeed, they circumvent two fundamental challenges of physics-based methods: i) the necessity to extensively sample the structure conformation space to estimate thermodynamic quantities and ii) the high computational cost of exploring the vast sequence space. Although evolutionary-based and physics-inspired machine learning methodologies are not systematically combined, they are highly synergistic in multiple aspects. Accordingly, the complementarity between evolutionary and physical modalities was successfully demonstrated for non-machine learning-based methods (see reviews by Marques et al. 2021 and Weinstein et al. 2020 [5,6]).

Here we will review recent achievements for both evolutionary-based and physics-inspired methods with emphasis on experimentally validated works. We will discuss the current limitations of both approaches and complementarities between them. Finally, we will review recent works combining both approaches and highlight possible future directions.

## Evolutionary-based design

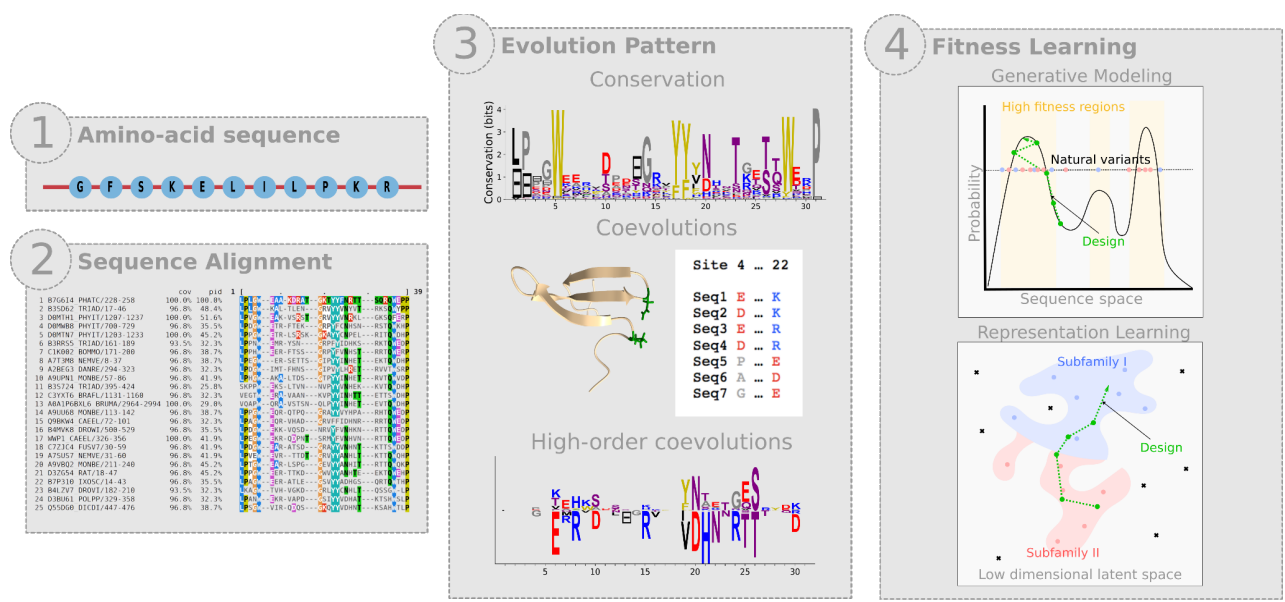

**Figure 1: Main steps of evolutionary-based design.** Starting from an amino-acid sequence, homologous sequences are retrieved from sequence databases and a multiple sequence alignment (MSA) is constructed. MSAs feature various evolutionary patterns including conservation, pairwise coevolution and high-order coevolution that reflect structural and functional constraints. Unsupervised machine learning distills fitness landscapes and representations from the evolutionary patterns, which in turn can be used for design.

Numerous design protocols involve modification of a preexisting natural protein towards improved or novel functional properties. Significant optimization of the target

property(ies) requires exploring sequences harboring many mutations from the wild type protein. However, it is estimated that up to 50% of single-point mutations are deleterious [7] for function, leading to exponentially decreasing success rates when mutating multiple sites. One solution is to restrict the search to mutations or combinations of mutations previously encountered, or likely to be encountered, throughout natural evolution of the protein. Evolutionary-based design consists in three main steps: (i) collection and alignment of a set of sequences homologous to the wild type protein that share similar structure and function, (ii) construction of a statistical/machine learning model that captures common patterns shared between these sequences such as conservation and coevolution and (iii) generation of artificial sequences that are distinct from natural ones but preserve the common patterns.

While early models such as Position-Specific Scoring Matrices (PSSM) solely focused on capturing site-specific amino acid frequencies, increasingly complex statistical models based on self-supervised machine learning have been developed. The Direct Coupling Analysis (DCA) method captures both single-site and pairwise correlations [8–10] arising from coevolution, allowing modeling of epistatic effects and drastic reduction of the search space. For instance, Russ et al. [11] designed with DCA hundreds of diverse Chorismate Mutase enzymes with native-like functionality, with a high success rate (~30%). Based on the entropy of the sequence distributions, they estimated that $10^{85}$ of all $10^{125}$ possible sequences with same length were potential design candidates based on a single site model, whereas only $10^{25}$ were suitable when including pairwise correlations.

Recent works investigated various neural network-based machine learning generative models (Box 1) [12–18], achieving successes in various enzyme and nanobody design tasks. These networks have more flexible probability distributions than PSSM or DCA models, allowing integration into the model of higher-order MSA statistics, such as co-occurrence of amino acid n-uplets. Importantly, some of these methods also learn a continuous, low-dimensional latent space representation of the protein. Unlike in the original sequence space - where single-point mutations may reduce probability and abolish function - small motions in latent space generally preserve the probability, and function. Thus, it is easier to navigate and to learn sequence/function associations in such "hole-free" landscapes. In [19], Biswas et al. leveraged the latent representation of a language model (see below) and small-scale experimental data to design improved GFP variants. Lian et al. [20] used a Variational Autoencoder (a representation-based generative model) to generate synthetic sequences of yeast $SH_3$ domains, and evaluated their ability to replace a native $SH_3$ domain *in vivo*. They found that most of the sequences with native-like functionality mapped to a local, convex region of the latent space.

Beyond direct generation of novel protein sequences, evolutionary models can also predict fitness-improving mutations [12,21], guide library design for large-scale screening experiments [22,23], or, conversely, infer fitness landscape from directed evolution experiments [24].

One drawback of such family-level models is that they do not generalize across protein families and as such can only be applied to protein families that include a large number of sequences. This is particularly problematic for proteins only conserved in eukaryotes. One possible avenue for overcoming these limitations are protein language models, as they can simultaneously model unrelated sets of protein sequences. Trained on large, unannotated databases of protein sequences such as UniClust [25] or BFD [26], protein language models (e.g., UniRep [14], ProGen [27], ESM-1b [28], ProVis [29], ProtTrans [30], ProteinBERT [31]) are train to reconstruct a sequence from a corrupted version with ~10-20% of the residues masked or randomly mutated, or to predict the next amino acid of a protein sequence given the previous ones. MSA context can also be provided (MSATransformer [32]). General understanding of protein chemistry emerged from masked language modeling, such as similarities between amino acids, secondary structure elements, or tertiary contacts. The models can then be further fine-tuned to account for specifics of protein families, even for families with low number of sequences and/or diversity [14,33]. Madani et al. [33] used the ProGen2 language model, fine-tuned and conditioned on natural lysozyme families to design novel lysozymes with native-like function (66/90 of the designs) and/or low sequence identity to natural proteins (<40%). They further reported that both pre-training and fine-tuning were important for accurate activity prediction. In [34], Hie et al. used the ESM-1b language model to propose single-point mutants of various antiviral antibodies with reduced reconstruction error. After experimental characterization and recombination of the best mutants, they found that the binding affinity could be improved for 4/7 of the tested antibodies.

---

**Box 1:** Generative models, autoregressive models and representations

**Generative models** are parametric probability distributions over a high-dimensional space of the form $P_\theta(s)$, such that $P_\theta(s) > 0$ for all S and $\sum_s P_\theta(s)=1$. The set of parameters theta are learned from the data by maximizing the average log-probability $¿\log P_\theta(s)>¿$. Informally, this amounts to assigning high probability to observed sequences and low elsewhere (**Fig1:4**, upper panel). After learning, new sequences are generated by drawing samples from the probability distribution.

**Autoregressive generate models** are a special class of generative models of the form $P_\theta(s)=P_\theta(s_1)P_\theta(s_2\vee s_1)P_\theta(s_3\vee s_1,s_2)\ldots$. The conditional distributions $P_\theta(s_j\vee s_{¿j})$ are parameterized by neural networks. Autoregressive models can be trained by maximizing the maximum likelihood as above, or, more efficiently, by masked modeling. The latter consists of masking (i.e. replacing by a placeholder) a random subset of positions (10-20%), and predicting their value from the remaining unmasked positions, via the conditional distributions. Informally, masked modeling enables learning of the interdependencies between the variables. After training, samples are generated by recursively sampling the next amino acid of a sequence given all the previous ones.

A **representation** is any deterministic mapping $R(S)$ from one high-dimensional space (e.g., the sequence space) to a continuous space, of typically lower dimension. Representations are especially useful for modeling proteins sequences when key properties $f(S)$ of the sequences (e.g. probability density, biological fitness, substrate specificity, phylogeny) can be well approximated as $f_R(R(S))$, where $f_R$ is a *continuous, smooth*

> function (**Fig1:4**, lower panel). This contrasts with the original sequence space, where small changes in the sequence can result in large changes of the properties (*e.g.*, a single mutation can abolish function). Representation can be learned either via unsupervised or supervised learning, by parameterizing the target function $f(S)$ (probability or biochemical property) as $f_R(R(S))$ where $R(S)$ is a trainable low-dimensional representation and $f_R$ is smooth and trainable.

## Physics-inspired methods

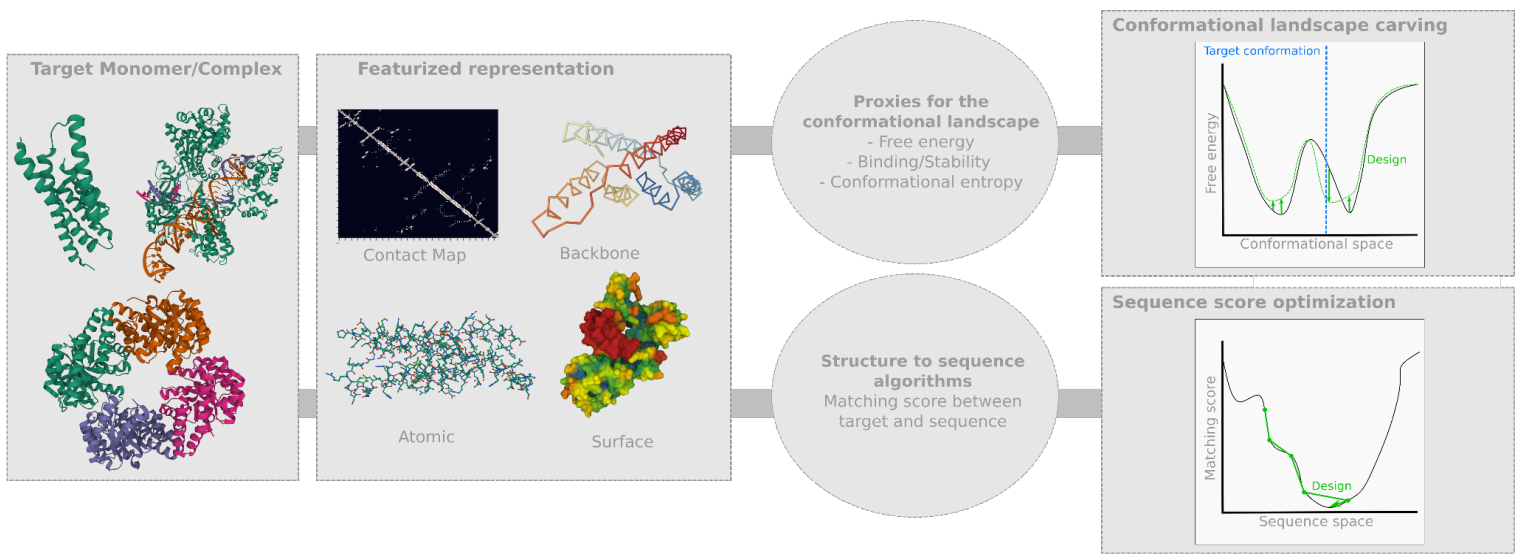

**Figure 2: Main steps of physics-inspired design**. Starting from a (partially or fully prescribed) target structure, one first builds a featurized representation suitable for relevant deep learning algorithms. Next, two complementary approaches are possible. First, sequence-to-structure prediction algorithms can be leveraged to build proxies for the free energy landscape. The latter is then used to design a sequence whose free energy minimum lies at the target conformation. Second, structure-to-sequence algorithms can be used to generate suitable sequences from structure. They rely on a matching score between target and sequence that can be optimized to find suitable sequences.

The protein design problem amounts to finding a sequence with low free energy in prescribed conformational state(s) (as a monomer structure, in complex with ligands). Classically, this is achieved by i) defining a tractable approximation of the free energy, and ii) minimizing it over the sequence. A common validation metric of design protocols is their sequence recovery rate: starting from the structure of a protein, its sequence-defined side chains are stripped, and a novel sequence is reconstructed from the remaining backbone atoms. The sequence recovery rate is the average sequence identity between the designed and original sequence: high values indicate that the structure-induced constraints on sequence are well recapitulated by the protocol.

Towards this end, a myriad of approximate physical force fields have been developed for protein design applications, including Rosetta [35,36] or FoldX [37]. These force fields are however heuristics that do not faithfully account for the underlying quantum dynamics and for the ability of the sequence to effectively fold into said structure from an unfolded state. Moreover, evaluating the free energy further involves a thorough exploration of the conformation state beyond the target states. Limitations of force-field based protein design include high computational cost (slow and inefficient Monte Carlo-based optimization), unsatisfactory sequence recovery rates from backbone structure (30-50%) and limited experimental success rate.

Recent progresses in machine learning-based structure prediction algorithms from sequence have opened the way to novel design paradigms. Deep learning models such as RaptorX [38], AlphaFold1 [39] or trRosetta [40] predict from multiple sequence alignments various geometric features of protein structures, such as contacts, backbone dihedral angles, pairwise distance matrices or angles between residue frames. Then, the model-computed negative log-probability of the features of a conformation defines an effective potential that can be minimized to fold the protein. By analogy with the Gibbs-Boltzmann ensemble, this effective potential can be interpreted as the free energy of the folded state. This approximation bypasses the requirement to extensively sample the conformation space for free energy estimation [37]. Fixed-backbone design protocols based on trRosetta [37], AlphaFold2 [41] and language models [42] were recently proposed and experimentally validated.  For binder design, binding free energy proxies can be similarly constructed: Gainza et al. [43,44] use a DL-computed matching score between molecular surface patches to design *de novo* mini-binding proteins.

What if the target conformation is partially or fully unspecified? Sequences with well-defined conformation can be similarly designed by minimizing the conformational entropy rather than the free energy. In these so-called hallucination protocols (initially introduced by *DeepDream* in the field of Computer Vision), the conformational entropy is quantified by the uncertainty of the structure prediction models. This is motivated here by the observation that low-confidence structure predictions often correspond to disordered regions of proteins [45]. Hallucination protocols based on trRosetta [46], AlphaFold2 and RoseTTAfold were recently proposed [47,48].

Following these approaches, several groups have successfully designed proteins with fully specified [49],[50] or de novo backbone structures and constructed folds around functional motifs [47,51]. A potential limitation is sequence diversity and amino acid compositional bias. Indeed, amino acids with versatile conformations (*e.g.* multiple side-chain torsion angle) and thus inherently uncertain structures are unfavored in such confidence-maximization protocols. Thus, these design protocols may not encompass the full diversity of sequences that can effectively adopt a target fold. Another weakness of these protocols is the potential existence of "adversarial" optima: sequences with highly confident predictions for one model but not another, *i.e.* that "trick" the network rather than solve the design problem [52].

Alternatively, one can try to directly predict suitable sequences for a given fold, the so-called inverse folding problem. Unlike the folding problem, many sequences can adopt the exact same fold and hence, a distribution of sequences should be constructed. One approach is to "thread" a sequence along the target backbone, calculate a compatibility score between the sequence and the backbone, and iteratively mutate it to improve its score. Using a 3D-convolutional neural network, Anand et al. [53] predict possible amino-acid substitutions and corresponding rotamer states from current structure, and iteratively mutate the protein to generate diverse sequences folding into a TIM-barrel. A second approach is to directly build a tractable distribution of sequences that can be easily sampled from: autoregressive generative models based on coarse-grained, graph-based representation of the protein backbone allow generation of a full sequence in a single pass [54–59]. These models achieve higher sequence recovery rates and are much less computationally intensive than force field-based approaches. They have been proposed for fixed-backbone monomer design [60] as well as multimers [61] and antibody design [62] [63]. Recently, Dauparas et al. [59] successfully designed various proteins using an autoregressive message-passing neural network.

Despite recent progress, these approaches still have some scope limitations: by construction, they are not suitable for modeling disordered proteins or segments. Fine-tuning of allosteric motion or of catalytic activity remains a major challenge, as these models are trained on static structures and are coarse-grained. Another source of concern is that these models are increasingly diverging from physics: for instance, AlphaFold implicitly assumes the presence of molecular cofactors, post-translational modifications or protein partners to properly fold a structure [64,65]. Thus, highly confident *in-silico* predictions may prove false in experimental conditions, and identification of model-derived matching scores to target physical properties is not always correct. For instance, some of the proteins designed by the trRosetta-based hallucination protocol [46], formed homo-oligomers or aggregates *in-vitro*, at odds with the monomeric *in-silico* prediction.

## Synergistic methods

Evolutionary-based and physics-inspired approaches are highly complementary for both coverage and scope, and have been used together extensively in classical protein design pipelines [5,6]. While physics-inspired models predict general biochemical properties (stability of the monomer and protein-ligand or protein-protein complexes), evolution-based methods learn various family-specific functional constraints in an agnostic fashion, including stability or catalytic activity, but also allosteric couplings [66] or specification of homo-oligomer state [67]. Thus, it is appealing to combine both methods for optimal success rate. This can be achieved in multiple ways.

First, the evolutionary model can be used to rapidly generate diverse sequence libraries, and then, candidate sequences would be prioritized based on the scores obtained from more computationally intensive physics-inspired models. Examples include the PROSS and FUNCLIB web servers, which use Rosetta and PSSM information to automatically redesign enzymes towards increased stability or modified catalytic activity [68,69]. Tran et al. and Das et al. designed, respectively, cell-penetrating and antimicrobial peptides using generative models combined with molecular dynamics [70] [71].

This simple approach is sufficient if a substantial fraction of evolutionary-designed sequences have satisfactory physical scores. Otherwise, multi-objective optimization/Monte Carlo sampling may be necessary to generate sequences that have both high evolutionary likelihood and physical score. The Rescue protocol redesign sequences by optimizing a weighted sum of the Rosetta energy and the evolutionary score estimated by a Potts model [72].

An open question is whether separate physical and evolutionary models are necessary. Instead, could one learn physical interactions from evolution, and reciprocally, to predict evolution from structure? A promising direction is to train joint models of MSA and structure, adapted from language and structure prediction models [73,74]. Other options could include fine-tuning of structure-based generative models using evolutionary information, or, conversely, regularization of evolutionary models using structural information.

Should we always restrict the search space for solving a given design problem to the vicinity of a specific protein family? Indeed, recent design approaches based on pan-family, transformer-based sequence generative models [60], diffusion-based structure generative models [75] and joint sequence/structure diffusion-based generative models [66] enable search beyond a specific family. The latter showed impressive successes *in-vitro* for monomer and binder design. If i) the design problem is well-defined in terms of target properties (*e.g.* thermostability, binding affinity to a target,...), ii) the target properties can be faithfully approximated *in-silico*, and iii) efficient exploration algorithms exist, then increasing the search space is a promising strategy. Otherwise, and especially if the problem is ill-defined (*e.g.* finding a "human-like" protein with improved thermostability and catalytic efficiency for gene therapy [76], redesigning a yeast $SH_3$ domain while preserving its cellular protein interactions [20]), then restricting the search space to the vicinity of a protein family is desirable.

# Summary and futures directions

In summary, evolutionary-based and physics-inspired approaches have undergone drastic improvement in the last years, enabling unprecedented success rates and democratization of protein design. However, important challenges remain.

Regarding evolutionary-based models, more efforts are required to develop models that can generalize among protein families, while maintaining computational tractability (for accessibility to the whole community), interpretability, and well-defined sampling protocols. Incorporation of known structural information as priors, e.g. using structure-aware transformer models such as EvoFormer [41] (where the structure is provided as template) is another interesting direction. Moreover, training protocols of evolutionary models should better account for phylogenetic relations between samples. In [77], Weinstein et al. argue that fitness prediction performances may plateau or decrease as model complexity increases without proper treatment of phylogeny - a phenomenon very recently observed with language models [78]. A conceptual and practical limitation is the entanglement of evolutionary constraints: current models do not have the ability to selectively discard specific constraints that are relevant *in-vivo* but not for engineered proteins (*e.g.* requirement to bind inhibitor protein(s), adequacy to specific cellular compartment,…). Conversely, proteins that are evolutionary-fitted at the family level may not be suitable for a specific biochemical task (*e.g.*, they may bind a related ligand, but distinct from the prescribed one).

Physics-inspired methods have made impressive progress for the design of monomers and oligomers with fully or partially-specified structure. In particular, spectacular improvements in success rates were reported for the *de novo* binder design problem [75,79], although subsequent *in vitro* directed evolution remains the norm [43,74]. Altogether, finding sequences that adopt a well-defined conformation is becoming easier. On the other hand, modeling of flexible regions such as immunoglobulin loops or peptides, catalytic sites (which require fine-grained description of intermediate catalytic states), or allosteric motions remains challenging. Another key challenge is to better characterize the *in vivo* behavior of these designed proteins (e.g., off-target distribution of protein binders, lifetime, immunogenicity). Future developments of ML-based Molecular Dynamics [80] and Neural Force fields docking algorithms such as the recent DiffDock [81] could open up the way to a better understanding of these functions. Extensions to other types of ligands, such as ions, nucleotides, or small organic molecules, are also important future directions.

While synergistic approaches have been widely successful for non-ML design protocols [5,6], most recent ML-based studies considered either of the approaches. Yet, evolutionary-based and physics-inspired modeling are highly complementary, and combining them through simple proposal/acceptance or joint optimization protocols should lead to higher success rates for challenging design problems. In the longer run, models predicting MSAs from structures, or, conversely, structure-based priors [82,83] for evolutionary models may allow overcoming the current limitations of each approach.

## Credit author statement

**C.M.** : Conceptualization, Investigation, Writing - Original Draft. **D.B.**, **S.C.**, **R.M.** : Conceptualization, Writing - Review & Editing, Supervision, Project administration, Funding acquisition. **J.T.** : Conceptualization, Investigation, Writing - Original Draft, Supervision, Funding acquisition.

## Conflict of interest statement

## Acknowledgements

J.T. was supported by the Edmond J. Safra Center for Bioinformatics at Tel Aviv University and from the Human Frontier Science Program (cross-disciplinary postdoctoral fellowship LT001058/2019-C). C.M. is recipient of a PhD funding from AMX program, Ecole Polytechnique and benefits from financial support from the Learning Planet Institute (LPI) through 'Ecole Doctorale Frontières de l'Innovation en Recherche et Education—Programme Bettencourt'.

integrates evolutionary constraints via Direct Coupling Analysis and physical energy scores via Rosetta. Their protocol achieved a higher sequence recovery rate (70%) than protocols that do not include evolutionary information or only single-site frequencies.